\documentclass[english]{article}
\usepackage{amsmath,bm,amssymb,soul,mathtools,mathptmx}
\DeclareMathAlphabet{\mathcal}{OMS}{cmsy}{m}{n}
\SetMathAlphabet{\mathcal}{bold}{OMS}{cmsy}{b}{n} 
\usepackage[T1]{fontenc}
\usepackage[letterpaper]{geometry}
\usepackage{float}
\usepackage{amsmath}
\usepackage{graphicx}
\usepackage{color}
\usepackage{epstopdf}
\usepackage{verbatim}
\usepackage[colorlinks, hypertexnames=false]{hyperref}
\hypersetup{linkcolor=blue, citecolor=red} 
\usepackage{adjustbox}
\usepackage{bm}
\usepackage{multirow}
\usepackage{array}
\usepackage{amsmath, amssymb, amsthm}

\usepackage{subfigure}
\usepackage{tikz}
\usepackage{algorithm}
\usepackage{algpseudocode}
\usepackage{enumerate}

\usepackage{amsfonts} 
\usepackage{color}
\usepackage{adjustbox}
\definecolor{black}{rgb}{0,0,0}

\definecolor{red}{rgb}{1,0,0}

\definecolor{blue}{rgb}{0,0,1}

\usepackage{multirow}

\DeclareOldFontCommand{\rm}{\normalfont\rmfamily}{\mathrm}
\DeclareOldFontCommand{\sf}{\normalfont\sffamily}{\mathsf}
\DeclareOldFontCommand{\tt}{\normalfont\ttfamily}{\mathtt}
\DeclareOldFontCommand{\bf}{\normalfont\bfseries}{\mathbf}
\DeclareOldFontCommand{\it}{\normalfont\itshape}{\mathit}
\DeclareOldFontCommand{\sl}{\normalfont\slshape}{\@nomath\sl}
\DeclareOldFontCommand{\sc}{\normalfont\scshape}{\@nomath\sc}
\DeclareRobustCommand*\cal{\@fontswitch\relax\mathcal}
\DeclareRobustCommand*\mit{\@fontswitch\relax\mathnormal}

\makeatother
\usepackage{mathtools}
\usepackage{esint}
\usepackage{cleveref}
\usepackage{natbib}
\usepackage{babel}
\usepackage{authblk}

\title{\textbf{A Lagrangian Conditional Gaussian Koopman Network for Data Assimilation and Prediction}}

\author[1]{Zhongrui Wang}
\author[2]{Chuanqi Chen}
\author[2]{Jin-Long Wu}
\author[1,*]{Nan Chen}
\affil[1]{Department of Mathematics, University of Wisconsin–Madison, Madison, WI 53706, USA}
\affil[2]{Department of Mechanical Engineering, University of Wisconsin–Madison, Madison, WI 53706, USA}
\affil[*]{\small{Corresponding author: Nan Chen, chennan@math.wisc.edu}}

\begin{document}
	\maketitle
	\begin{abstract}
        Lagrangian data assimilation seeks to recover hidden Eulerian flow fields from sparse and indirect observations of moving tracers. This problem is fundamentally challenging because of the nonlinear coupling between tracer trajectories and the underlying flow, rendering posterior inference computationally intractable for realistic, high-dimensional systems. In this work, we develop a Lagrangian conditional Gaussian Koopman network (LaCGKN), a structure-preserving and data-driven framework for joint data assimilation and prediction from Lagrangian observations. LaCGKN embeds the Eulerian flow dynamics into a low-dimensional latent space governed by a nonlinear stochastic system with conditional Gaussian structures, enabling analytic posterior updates without ensemble forecasting. Different from existing conditional Gaussian Koopman formulations that rely on direct Eulerian observations, the Lagrangian setting introduces additional constraints on the latent representation, which must simultaneously encode the flow dynamics and mediate nonlinear tracer-flow interactions. To address these challenges, the LaCGKN incorporates three key architectural components: (i) tracer homogenization to enforce permutation equivariance and enable generalization across varying numbers of tracers; (ii) Fourier-based positional encoding to capture spatial dependence and reconstruct local flow features at moving tracer locations; and (iii) an SVD-inspired low-rank parameterization of the latent transition operator, which reduces parameter complexity while preserving expressive capacity. An application to a two-layer quasi-geostrophic flow with surface tracer observations demonstrates that LaCGKN achieves accurate and efficient Lagrangian data assimilation and prediction, without reliance on ensemble methods or the governing physical model. These results establish the LaCGKN as a unified and computationally tractable alternative to both traditional model-based approaches and purely black-box data-driven methods.
    \end{abstract}

%
\section{Introduction}

Lagrangian data consist of trajectories of moving tracers, such as drifters and floats, that follow particle motions within a flow. By recording the time-evolving positions of these tracers, Lagrangian observations provide a direct window into the transport properties of the underlying flow field and are widely used to characterize large-scale circulation patterns, coherent structures, and mixing processes \citep{businger_balloons_1996, gould_argo_2004,lumpkin_advances_2017, centurioni_global_2017}. While valuable information can be extracted from Lagrangian trajectories through direct analysis, a more systematic and powerful approach is to integrate these observations with dynamical or statistical models through Lagrangian data assimilation. In the Lagrangian setting, the data assimilation objective is to estimate the underlying flow state using observations of tracer trajectories \citep{mariano_lagrangian_2002, ide_lagrangian_2002}. The importance of Lagrangian data assimilation in both geophysical research and operational forecasting continues to grow, driven by the rapid expansion of global Lagrangian observing systems \citep{legler_current_2015, mcphaden_tropical_2023}. However, a fundamental challenge arises from the nature of the observation process: Lagrangian measurements are obtained by evaluating the local flow velocity at tracer positions that evolve in time, resulting in a highly nonlinear observation operator \citep{ide_lagrangian_2002, kuznetsov_method_2003}. When combined with the strongly nonlinear and often turbulent dynamics of geophysical flows, this intrinsic nonlinearity renders Lagrangian data assimilation particularly challenging. As a consequence, analytic expressions for the posterior distribution are generally intractable for realistic, high-dimensional systems.

Early approaches to Lagrangian data assimilation addressed nonlinearity by augmenting Eulerian flow models with explicit tracer dynamics and applying Kalman-filter-based methods. Representative examples include the use of the extended Kalman filter (EKF) to assimilate tracer trajectories \citep{ide_lagrangian_2002}, the optimal interpolation-based schemes that treat finite-difference Lagrangian velocities as surrogate observations with constant gain \citep{molcard_assimilation_2003}, and applying the local ensemble transform Kalman filter (LETKF) to drifters \citep{sun_lagrangian_2019, chen_efficient_2022}. While these approaches remain attractive for operational applications, their reliance on Gaussian assumptions and linear correction steps can sometimes limit accuracy in strongly nonlinear regimes. To better represent nonlinearity and non-Gaussianity, particle filters \citep[PF;][]{van_leeuwen_particle_2019} and Markov chain Monte Carlo (MCMC) smoothers \citep{apte_bayesian_2008} have been explored, but their computational cost poses severe challenges for high-dimensional systems. Hybrid strategies attempt to balance the trade-off between approximation accuracy and computational cost, for example, by combining particle filtering for nonlinear tracer dynamics with ensemble Kalman filtering for near-Gaussian flow estimation \citep{slivinski_hybrid_2015}, or by exploiting special structures to account for flow nonlinearity \citep{wang_nonlinear_2025}. Despite these advances, designing Lagrangian data assimilation methods that simultaneously achieve high accuracy and computational efficiency remains an open challenge.

Traditional data assimilation methods are predominantly deductive (model-based), derived from first principles through Bayes’ rule. Posterior inference is carried out explicitly via variational optimization, such as 3D/4D-Variational methods \citep{lorenc_analysis_1986, lorenc_modelling_2003}, or Monte Carlo-based filtering approaches, such as the ensemble Kalman filter \citep[EnKF;][]{evensen_sequential_1994}. To remain computationally feasible for high-dimensional systems, these methods typically impose strong structural assumptions, most notably Gaussian error statistics and linearized dynamics or observation operators.
Driven by advances in machine learning, inductive (data-driven) approaches to data assimilation have gained increasing attention \citep{cheng_machine_2023, bach_learning_2025}. Developments in scientific machine learning (SciML) have enabled flexible surrogate models for complex dynamical systems, including recurrent neural networks \citep{schuster_bidirectional_1997, hochreiter_long_1997, gauthier_next_2021}, neural ODEs \citep{chen_neural_2019}, physics-informed neural networks \citep{raissi_physics-informed_2019}, neural operators \citep{lu_learning_2021, li_fourier_2021, chen_neural_2025}, and generative models for stochastic systems \citep{du_conditional_2024, dong_data-driven_2025, stamatelopoulos_can_2025}. When combined with traditional filters and smoothers, SciML can enhance data assimilation by providing efficient forecast models \citep{penny_integrating_2022}, correcting model errors \citep{farchi_online_2023}, or learning components of the analysis map \citep{bach_learning_2025}. More recently, hybrid approaches that preserve analytically tractable nonlinear structures, such as conditional Gaussian neural stochastic differential equations, have been proposed to balance expressiveness with efficient posterior inference \citep{chen_cgnsde_2024}. Furthermore, fully data-driven methods attempt to learn the analysis or posterior mapping directly \citep{bocquet_accurate_2024}. Other examples include ensemble transport \citep{spantini_coupling_2022}, generative data assimilation \citep{qu_deep_2024, li_probabilistic_2025, martin_generative_2025, yang_generative_2025}, and approaches that jointly learn forecast and analysis maps \citep{fablet_learning_2021, boudier_data_2023}. Collectively, these developments span a spectrum from deductive to inductive, highlighting an ongoing exploration of the trade-off among physical interpretability, statistical rigor, and computational efficiency.

However, most data-driven data assimilation methods have been developed for Eulerian observations on fixed spatial grids. In contrast, the use of machine learning for Lagrangian data assimilation remains relatively limited, with only a few recent examples combining neural operators with generative models \citep{asefi_generative_2025} or multimodal contrastive learning \citep{baptista_mathematical_2025}. This disparity is notable, as machine learning is particularly well suited to addressing the strong nonlinearity inherent in Lagrangian observations and therefore has substantial potential to improve both accuracy and computational efficiency.
A related development is the conditional Gaussian Koopman network \citep[CGKN;][]{chen_cgkn_2025, chen_modeling_2025}, which introduces a data-driven dynamical model embedded within a conditional Gaussian nonlinear stochastic system that admits analytic posterior updates and avoids various tunings often required by ensemble-based methods. By incorporating principled filtering formulas directly into the model architecture as an inductive bias, CGKN provides a structured interface between SciML and data assimilation. However, existing CGKN formulations are restricted to systems with direct, partial Eulerian observations, and do not address the fundamental challenges posed by Lagrangian data.

In this work, we develop a Lagrangian conditional Gaussian Koopman network (LaCGKN) for estimating the underlying flow field from tracer trajectories. Beyond the intrinsic nonlinearity of Lagrangian data assimilation, this setting introduces challenges that are qualitatively distinct from those encountered in CGKN with Eulerian observations. In particular, the latent representation must simultaneously support an efficient and accurate encoding of the Eulerian flow dynamics and a nonlinear yet structured observation mechanism that governs tracer motion. The latter plays a central role in Lagrangian data assimilation, as it determines how information carried by moving tracers propagates to the unobserved flow field, thereby imposing additional constraints on the latent embedding that are absent in Eulerian settings.
To address these challenges, we construct a purely data-driven surrogate that embeds the Eulerian flow into a low-dimensional latent space while modeling tracer dynamics as a nonlinear function of tracer positions driven by the latent state. This design preserves a nonlinear yet conditional Gaussian structure, enabling analytic filtering in latent space without ensemble forecasting of the physical model. The resulting LaCGKN incorporates three key architectural components:
(i) tracer homogenization, which enforces tracer permutation equivariance and enables generalization across varying numbers of tracers;
(ii) Fourier-based positional encoding, which captures rich spatial dependence and facilitates reconstruction of local flow features at moving tracer positions; and
(iii) an SVD-inspired low-rank parameterization, which maintains expressive capacity at high latent dimensions while controlling computational cost.
Together, these design choices enable accurate and efficient Lagrangian data assimilation and prediction, offering a unified alternative to both traditional model-based approaches and purely black-box data-driven methods.


The rest of the paper is organized as follows. Section 2 introduces the formulation of Lagrangian data assimilation and discrete-time CGKN, and then formalizes the LaCGKN for data assimilation and prediction, highlighting the key architectural design choices. Section 3 demonstrates the proposed method with an application to a two-layer quasi-geostrophic flow with surface tracer observations. For prediction, LaCGKN is compared to a deep-learning baseline that uses a deep neural
network (DNN) for tracer prediction and a convolutional neural network (CNN) for flow prediction, as well as a trivial persistence baseline. For data assimilation, LaCGKN is compared to EnKF, OI, and a naive climatology baseline. Section 4 concludes the paper with discussions.

\section{Methodology}
\subsection{Lagrangian data assimilation}
Lagrangian data assimilation aims at assimilating Lagrangian observations that are often driven by a flow described under Eulerian coordinates. To illuminate the idea, we focus on a canonical setup: Lagrangian observations of passive tracer positions driven by an Eulerian flow, described as
\begin{subequations}\label{eq:tracerflow}
\begin{align}
\frac{\mathrm{d}\mathbf{x}_i}{\mathrm{d}t} &= \mathbf{v}(\mathbf{x}_i, t) + \boldsymbol{\Sigma}_{\mathbf{x}_i}\dot{\mathbf{W}_i},  \quad \ i = 1, \ldots, I  \label{eq:tracer} \\
\frac{\partial \mathbf{v}}{\partial t} &= \mathcal{F}[\mathbf{v}] + \boldsymbol{\Sigma}_{\mathbf{v}}\dot{\mathbf{W}_\mathbf{v}},\label{eq:flow}
\end{align}
\end{subequations}
where $\mathbf{x}_i(t) \in \Omega\subset \mathbb{R}^d$ is the observed position of the $i$th tracer, and $\mathbf{v}(\mathbf{x}, t)\in \mathbb{R}^d$ is the flow velocity field with $(\mathbf{x},t) \in \Omega \times (0,T]$. The flow operator $\mathcal{F}$ is nonlinear in general. The system is stochastic with the Gaussian white noise $\dot{\mathbf{W}}$ and the noise strength matrix $\boldsymbol{\Sigma}$, representing the uncertainty due to the unresolved details. This formulation can be generalized to arbitrary Lagrangian observations by modifying (\ref{eq:tracer}) describing the Lagrangian observation with dependence on the tracer position. This canonical setup already reveals a key challenge in Lagrangian data assimilation. That is, the Lagrangian observational process is nonlinear, because the composition $\mathbf{v}(\mathbf{x}_i(t), t)$ is generally nonlinear in $\mathbf{x}_i(t)$.

In practice, a discrete version of (\ref{eq:tracerflow}) is considered as follows:

\begin{subequations}\label{eq:tracerflow_discrete}
\begin{align}
\mathbf{x}_i^{n+1}
&= \mathcal{T}_h\big(\mathbf{x}_i^{n},\mathbf{v}^{n}\big)
  + \boldsymbol{\Sigma}_{\mathbf{x}_i}\,\Delta \mathbf{W}_i^{\,n},
\quad i=1,\ldots,I, \label{eq:tracer_discrete} \\[4pt]
\mathbf{v}^{n+1}
&= \mathcal{F}_h\big(\mathbf{v}^{n}\big)
  + \boldsymbol{\Sigma}_{\mathbf{v}}\,\Delta \mathbf{W}_{\mathbf{v}}^{\,n},
\label{eq:flow_discrete}
\end{align}
\end{subequations}
where $\mathbf{x}_i^{n}:=\mathbf{x}_i({t_n}) \in \Omega$ and $\mathbf{v}^{n}:=\mathbf{v}(\cdot, t_n)|_{\Omega_{\mathbf{x}}^{(M)}} \in \mathbb{R}^{dM}$ are the corresponding state variables evaluated at discrete time $t_n$ with time interval $t_{n+1}-t_n=\Delta t$ and an $M$-point discrete spatial domain $\Omega_{\mathbf{x}}^{(M)}$. The $\mathcal{T}_h$ and $\mathcal{F}_h$ are the solution maps of tracer and flow, respectively, from $t_n$ to $t_{n+1}$. The tracer operator $\mathcal{T}_h$ typically involves an interpolation or integral operator (e.g., Fourier transform) that evaluates $\mathbf{v}(\mathbf{x}_i^n, t_n)$ at a local tracer position $\mathbf{x}_i^n$.

The Lagrangian filtering problem defined in the discrete setup (\ref{eq:tracerflow_discrete}) seeks to find the posterior distribution of the unobserved flow states $\mathbf{v}^n$ given past data of tracer observations $\{\mathbf{x}_i^n\}_{i=1, s=0}^{I, n}$, that is
\begin{equation}\label{eq:posterior}
    p(\mathbf{v}^n |\{\mathbf{x}_i^s\}_{i=1, s=0}^{I, n}).
\end{equation}
Since the solution maps $\mathcal{T}_h$ and $\mathcal{F}_h$ are typically nonlinear, an analytic solution of the posterior distribution is almost intractable for high-dimensional systems in practice, making Lagrangian data assimilation a challenging problem. 

\subsection{Discrete-time CGKN}
The conditional Gaussian Koopman network \cite[CGKN;][]{chen_cgkn_2025,chen_modeling_2025} is a unified deep learning framework that allows efficient data assimilation and state forecast. It transforms the original dynamical system into one with a conditional Gaussian structure, where the posterior distribution of embedded unobserved variables can be solved via analytic formulae. The transformed dynamical system serves as a surrogate model, enabling efficient data assimilation and state forecast. The assimilated and predicted states are then mapped back to the original state space to obtain the final results. 

Consider a partially observed discrete-time dynamical system in the general form of  
\begin{subequations}\label{eq:discrete_system}
\begin{align}
    \mathbf{u}_1^{n+1} &= \mathcal{G}_1\big( \mathbf{u}_1^{n}, \mathbf{u}_2^{n}\big), \\
    \mathbf{u}_2^{n+1} &= \mathcal{G}_2\big( \mathbf{u}_1^{n}, \mathbf{u}_2^{n}\big),
\end{align}
\end{subequations}
where $\mathbf{u}_1 \in \mathbb{R}^{d_1}$ and $\mathbf{u}_2 \in \mathbb{R}^{d_2}$ are vectors representing the observed states and unobserved states, respectively. The state transition operators $\mathcal{G}_1$ and $\mathcal{G}_2$ map the solutions from $t_n$ to $t_{n+1}$, with time interval $\Delta t=t_{n+1}-t_{n}$. 

For the discrete dynamical system (\ref{eq:discrete_system}), CGKN transforms it into a surrogate model of the following form:
\begin{subequations}\label{eq:DiscreteCGKN}
\begin{align}
\mathbf{u}_1^{n+1} &= \mathbf{F}_1\big(\mathbf{u}_1^{n}\big) + \mathbf{G}_1\big(\mathbf{u}_1^{n}\big)\mathbf{z}^{n} + \boldsymbol{\Sigma}_1\boldsymbol{\epsilon}_1^{n},\label{eq:cgkn_u1}\\
\mathbf{z}^{n+1} &= \mathbf{F}_2\big(\mathbf{u}_1^{n}\big) + \mathbf{G}_2\big(\mathbf{u}_1^{n}\big)\mathbf{z}^{n} + \boldsymbol{\Sigma}_2\boldsymbol{\epsilon}_2^{n},\label{eq:cgkn_z}
\end{align}
\end{subequations}
where $\mathbf{z} \in \mathbb{R}^{d_{\mathbf{z}}}$ are latent embeddings of the original unobserved states $\mathbf{u}_2$. The transformation between latent states and original states is realized through $\mathbf{z} = \mathcal{E}(\mathbf{u}_2)$ and $\mathbf{u}_2=\mathcal{D}(\mathbf{z})$, where $\mathcal{E}:\mathbb{R}^{d_2}\mapsto\mathbb{R}^{d_{\mathbf{z}}}$ is the encoder and $\mathcal{D}:\mathbb{R}^{d_{\mathbf{z}}} \mapsto \mathbb{R}^{d_2}$ is the decoder. The coefficients $\mathbf{F}_1$, $\mathbf{G}_1$, $\mathbf{F}_2$, and $\mathbf{G}_2$ are functions of the observed states $\mathbf{u}_1$. The terms $\boldsymbol{\Sigma}_1 \boldsymbol{\epsilon}_1$ and $\boldsymbol{\Sigma}_2 \boldsymbol{\epsilon}_2$ are Gaussian white noise multiplied by the noise strength matrices. In practice, the encoder $\mathcal{E}$, the decoder $\mathcal{D}$, and the coefficient functions $\mathbf{F}_1$, $\mathbf{G}_1$, $\mathbf{F}_2$, $\mathbf{G}_2$ are parameterized by neural networks. The noise strength matrix $\boldsymbol{\Sigma}_1$ is estimated by the one-step residual error of the observed states $\mathbf{u}_1$ using a trained CGKN without DA loss. The noise strength matrix $\boldsymbol{\Sigma}_2$ can be set manually or as a training parameter. 

The surrogate model (\ref{eq:DiscreteCGKN}) has a notable conditional Gaussian structure that allows for efficient data assimilation. Specifically, because $\mathbf{F}_1$, $\mathbf{G}_1$, $\mathbf{F}_2$, and $\mathbf{G}_2$ are only functions of $\mathbf{u}_1$, $\mathbf{z}$ is conditionally linear given an observed trajectory of $\mathbf{u}_1$. As a result, the conditional distribution $p(\mathbf{z}^{n}|\{\mathbf{u}_1^s\}^n_{s=0})=\mathcal{N}(\boldsymbol{\mu}_{\mathbf{z}}^{n}, \mathbf{R}_{\mathbf{z}}^{n})$ is Gaussian with the mean and covariance solvable by the analytic formulae:
\begin{equation}
\label{eq:CGFilter}
    \begin{aligned}
    \boldsymbol{\mu}_{\mathbf{z}}^{n+1} &= \mathbf{F}_{2}^n + \mathbf{G}_2^{n}\boldsymbol{\mu}_{\mathbf{z}}^{n} + \mathbf{K}^{n} \big(\mathbf{u}_1^{n+1} - \mathbf{F}_{1}^n - \mathbf{G}^n_1\boldsymbol{\mu}_{\mathbf{z}}^{n}\big)\\
    \mathbf{R}_{\mathbf{z}}^{n+1} &= \mathbf{G}_2^{n}\mathbf{R}_{\mathbf{z}}^{n}\mathbf{G}_2^{n\top} + \boldsymbol{\Sigma}_2\boldsymbol{\Sigma}_2^\top - \mathbf{K}^{n}\mathbf{G}^n_1\mathbf{R}_{\mathbf{z}}^{n}\mathbf{G}_2^{n\top}
    \end{aligned}
\end{equation}
where
\begin{equation}
    \mathbf{K}^{n} = \mathbf{G}_2^{n} \mathbf{R}_{\mathbf{z}}^{n}\mathbf{G}_1^{n\top}\big(\boldsymbol{\Sigma}_1\boldsymbol{\Sigma}_1^\top + \mathbf{G}^n_1\mathbf{R}_{\mathbf{z}}^{n}\mathbf{G}_1^{n\top}\big)^{-1}
\end{equation} 
and $\mathbf{F}^n_i:= \mathbf{F}_i\big(\mathbf{u}_1^{n}\big), \mathbf{G}^n_i := \mathbf{G}_i\big(\mathbf{u}_1^{n}\big)$, for $i=1,2$. The closed analytic formulae allow exact and accurate data assimilation solution and avoid empirical tunings as in many ensemble methods. The posterior mean of the latent states is transformed back to the original states through the decoder $\boldsymbol{\mu}_2^{n+1} = \mathcal{D}(\boldsymbol{\mu}_{\mathbf{z}}^{n+1})$. The posterior covariance $\mathbf{R}_2^{n}$ of the unobserved states can be calculated through residual analysis, a post-process of CGKN to quantify the uncertainty of the posterior mean. 

\subsection{LaCGKN}\label{sec:LaCGKN}
\begin{figure}
    \centering
    \includegraphics[width=1\linewidth]{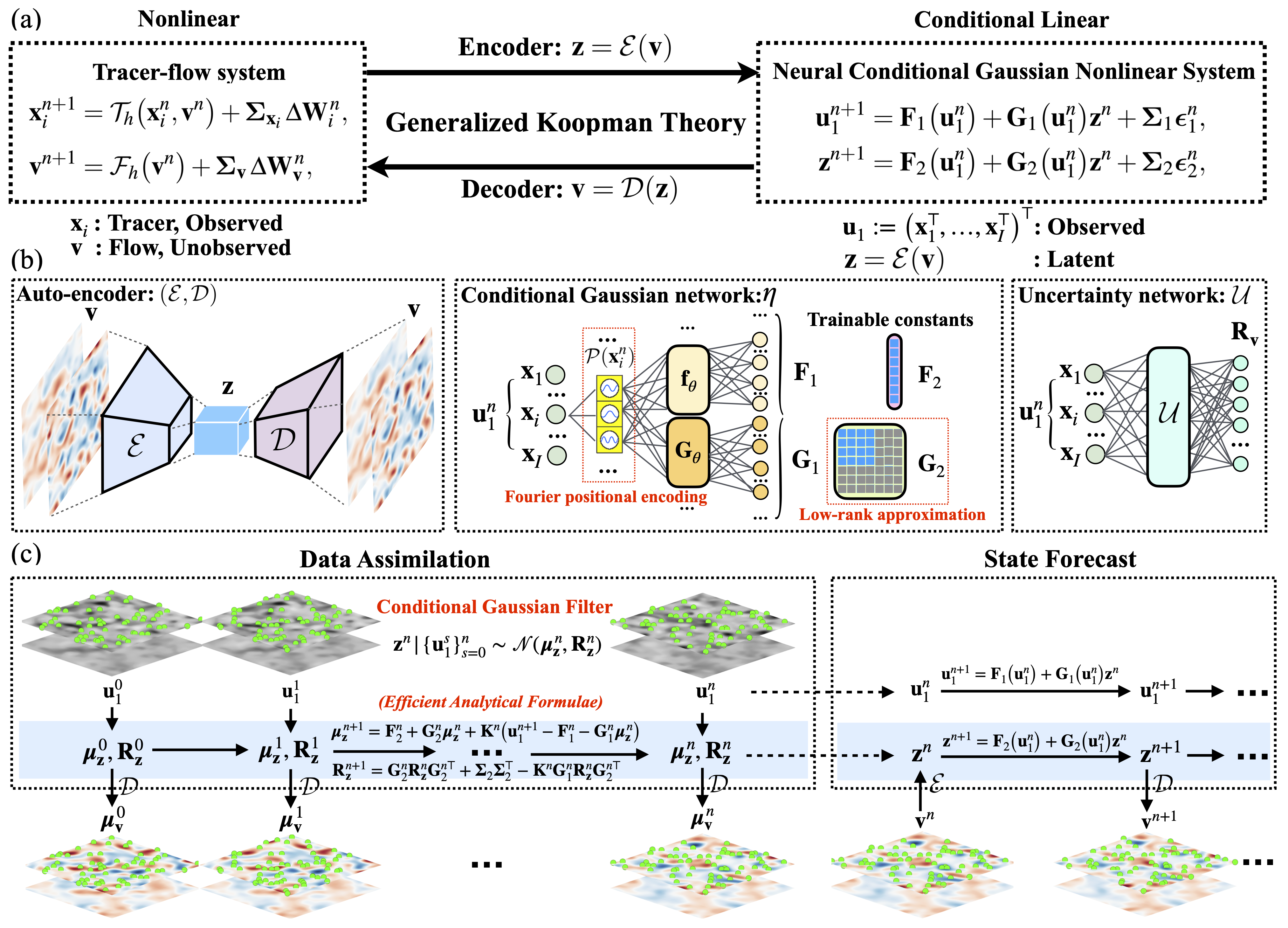}
    \caption{Overview of the Lagrangian conditional Gaussian Koopman network (LaCGKN). (a) The nonlinear tracer-flow system is mapped to a neural conditional Gaussian nonlinear system (Neural CGNS) by encoding the unobserved flow and mapped back by decoding the latent variables. (b) LaCGKN consists of an auto-encoder $(\mathcal{E},\mathcal{D})$ and conditional Gaussian network (CGN) $\eta$. The CGN outputs the coefficients $\mathbf{F}_1, \mathbf{G}_1, \mathbf{F}_2,\mathbf{G}_2$ of the Neural CGNS. The parameterization of $\mathbf{F}_1$ and $\mathbf{G}_1$ incorporates tracer homogenization and Fourier positional encoding, while $\mathbf{G}_2$ is represented by an SVD-inspired low-rank approximation. An auxiliary uncertainty network $\mathcal{U}$ is used to estimate the posterior standard deviation of the flow. (c) LaCGKN performs efficient data assimilation and prediction in latent space using the conditional Gaussian filter.}  
    \label{fig:LaCGKN_overview}
\end{figure}

\paragraph{LaCGKN surrogate.}
In the discrete tracer-flow system \eqref{eq:tracerflow_discrete}, Lagrangian tracer positions are observed and the Eulerian flow field is to be recovered. This leads to a natural definition of the observed and unobserved variables in the discrete-time CGKN:
\begin{equation}\label{eq:u1u2_LaCGKN}
    \mathbf{u}_1^{n}
    := \big(\mathbf{x}_1^{n\top},\ldots,\mathbf{x}_I^{n\top}\big)^\top \in \mathbb{R}^{dI},
    \qquad
    \mathbf{u}_2^{n}
    := \big(\mathbf{v}_1^{n\top},\ldots,\mathbf{v}_M^{n\top}\big)^\top \in \mathbb{R}^{dM},
\end{equation}
where $\mathbf{u}_1$ collects all the Lagrangian tracer positions and $\mathbf{u}_2$ contains the Eulerian flow states over discrete spatial domain $\Omega_{\mathbf{x}}^{(M)}$. The resulting LaCGKN surrogate model takes the form 
\begin{subequations}\label{eq:LaCGKN}
\begin{align}
\mathbf{u}_1^{n+1} &= \mathbf{F}_1\big(\mathbf{u}_1^{n}\big) + \mathbf{G}_1\big(\mathbf{u}_1^{n}\big)\mathbf{z}^{n} + \boldsymbol{\Sigma}_1\boldsymbol{\epsilon}_1^{n},\label{eq:LaCGKN_u1}\\
\mathbf{z}^{n+1} &= \mathbf{F}_2 + \mathbf{G}_2\mathbf{z}^{n} + \boldsymbol{\Sigma}_2\boldsymbol{\epsilon}_2^{n},\label{eq:LaCGKN_z}
\end{align}
\end{subequations}
which coincides with the general discrete-time CGKN formulation \eqref{eq:DiscreteCGKN}, except that $\mathbf{F}_2$ and $\mathbf{G}_2$ are now trainable constants instead of functions of $\mathbf{u}_1^n$. This simplification leverages the physical assumption that passive tracers do not influence the underlying flow. The Eulerian flow $\mathbf{u}_2^{n}$ is mapped into a low-dimensional latent embedding $\mathbf{z}^{n}$ via an encoder $\mathcal{E}$, and mapped back to physical states via a decoder $\mathcal{D}$. This flow embedding aligns with modern applied Koopman theory based on deep learning \citep{lusch_deep_2018}, but with a key distinction: the embedded flow $\mathbf{z}$ is coupled to the nonlinear tracer dynamics through \eqref{eq:LaCGKN_u1}. 
Together, the linear latent flow dynamics (\ref{eq:LaCGKN_z}) and the transformed tracer dynamics \eqref{eq:LaCGKN_u1} consist of the LaCGKN surrogate model. An overview of the LaCGKN is shown in Figure \ref{fig:LaCGKN_overview}.

\paragraph{LaCGKN for data assimilation and state forecast.}
LaCGKN performs efficient data assimilation and state forecasting in latent space. For data assimilation, the posterior mean and covariance of the latent state are updated analytically using the conditional Gaussian filter \eqref{eq:CGFilter}, which avoids ensemble propagation and empirical tunings. The posterior mean of the physical flow is then obtained by decoding the latent posterior mean via the decoder $\mathcal{D}$. While the posterior uncertainty of the physical flow does not admit a closed-form expression because of the nonlinear decoding, an auxiliary uncertainty network $\mathcal{U}$ can be employed as a post-processing step to estimate the posterior covariance in physical space. The network $\mathcal{U}$ takes the observed state $\mathbf{u}_1$ as input and predicts the residual $\|\mathbf{u}_2^\star - \boldsymbol{\mu}_2\|$, where 
$\mathbf{u}_2^\star$ denotes the ground truth and $\boldsymbol{\mu}_2 = \mathcal{D}(\boldsymbol{\mu}_{\mathbf{z}})$ is the decoded posterior mean. Under a Gaussian assumption, this residual provides a maximum-likelihood estimate of the posterior standard deviation of the physical flow. For state forecasting, the tracer state and latent flow are advanced using the LaCGKN surrogate model \eqref{eq:LaCGKN}, and the physical flow is recovered by decoding the predicted latent flow. Since both filtering and forecasting operate in a typically low-dimensional latent space, LaCGKN is highly efficient for high-dimensional turbulent systems.

It is worth highlighting that the Lagrangian observations (i.e., tracer positions) are indirect observations of the hidden system states (i.e., flow fields), and are coupled with the hidden states through a nonlinear dynamical process. This differs from previous applications of CGKN, where the observed variables are direct yet partial observations of the same underlying system. The Lagrangian data assimilation is thus a more challenging yet practical problem for CGKN, where not only the nonlinear flow dynamics needs to be well approximated by a latent linear dynamics, as in standard Koopman theory, but also the nonlinear tracer dynamics should be well approximated by (\ref{eq:cgkn_u1}). The latter is crucial to data assimilation, as it captures the information propagation from observed states to unobserved states. This imposes additional requirements and regularizations for latent embedding $\mathbf{z}$. Furthermore, when the modeling system is high-dimensional and highly turbulent, as is common in geophysical applications, a relatively large latent dimension may be needed, thus increasing the computational cost of CGKN. In this work, we address these challenges by introducing the following innovative design choices.

\paragraph{(a) Homogenization over tracers.} Lagrangian tracers can often be assumed to be homogeneous and exchangeable. This is already implied in tracer equation (\ref{eq:tracer}), where different tracers obey the same dynamical law. We therefore construct the LaCGKN observation operators $\mathbf{F}_1$ and $\mathbf{G}_1$ by applying the same neural networks to each tracer position independently, namely,
\begin{subequations}\label{eq:homo_tracers}
\begin{align}
\mathbf{F}_1\big(\mathbf{u}_1^{n}\big)
&:= \big(\mathbf{f}_\theta(\mathbf{x}_1^n)^\top,\ \ldots,\ \mathbf{f}_\theta(\mathbf{x}_I^n)^\top\big)^\top,\\
\mathbf{G}_1\big(\mathbf{u}_1^{n}\big)
&:= \big(\mathbf{G}_\theta(\mathbf{x}_1^n)^\top,\ \ldots,\ \mathbf{G}_\theta(\mathbf{x}_I^n)^\top\big)^\top,
\end{align}
\end{subequations}
where $\mathbf{f}_\theta:\mathbb{R}^d\to\mathbb{R}^{d}$ and $\mathbf{G}_\theta:\mathbb{R}^d\to\mathbb{R}^{d\times d_{\mathbf{z}}}$ are neural networks with shared parameters across all tracers. The tracer update then takes the form
\begin{equation}
    \mathbf{x}_i^{n+1}
    = \mathbf{f}_\theta(\mathbf{x}_i^n) 
      + \mathbf{G}_\theta(\mathbf{x}_i^n)\,\mathbf{z}^{n}
      + \text{noise}, \qquad i=1,\ldots,I.
\end{equation}
This design ensures that the surrogate dynamics is permutation-equivariant with respect to tracers, and it allows the same LaCGKN to be trained once and then applied to an arbitrary number of tracers in the test stage. In implementation, the invariance to tracer permutation is further encouraged by randomly subsampling different tracer subsets during training, while prediction and filtering always use the same set of shared networks $\mathbf{f}_\theta$ and $\mathbf{G}_\theta$. Compared to using large neural networks that take all tracer positions $\mathbf{u}_1^{n}$ as input to parameterize $\mathbf{F}_1$ and $\mathbf{G}_1$, tracer homogenization greatly regularizes the model, simplifies model development, and improves computational efficiency. 

\paragraph{(b) Fourier-based positional encoding.}
To accurately reconstruct the local value of the Eulerian flow field at moving tracer locations, $\mathbf{G}_\theta(\mathbf{x}_i^n)$ must learn a rich and nonlinear dependence on position. We therefore adopt Fourier-based positional encoding for tracer locations, which has been shown to significantly improve the ability of neural networks to represent high-frequency spatial variations \citep{tancik_fourier_2020, mildenhall_nerf_2020}. For a tracer at position $\mathbf{x}_i^n = (x_{i,1}^n,\ldots,x_{i,d}^n)$ on a periodic domain, a Fourier positional feature can be constructed by augmenting $\mathbf{x}_i^n$ with a sequence of trigonometric functions:
\begin{equation}\label{eq:FourierEncoding}
    \mathcal{P}(\mathbf{x}_i^n)
    = \Big[
        x_{i,1}^n,\ldots,x_{i,d}^n,\ 
        \big\{\sin(2^{k}\pi x_{i,j}^n),\cos(2^{k}\pi x_{i,j}^n)\big\}_{j=1,k=0}^{d,K-1}
    \Big],
\end{equation}
which is then fed into the networks $\mathbf{f}_\theta$ and $\mathbf{G}_\theta$. This encoding also admits physical heuristics from the tracer equation. Recall that the solution operator  $\mathcal{T}_h\big(\mathbf{x}_i^{n}, \mathbf{v}^{n}\big)$ in  (\ref{eq:tracer_discrete}) typically involves evaluating the Eulerian velocity field at tracer locations $\mathbf{v}(\mathbf{x}_i^n, t_n)$ through Fourier transforms, where Fourier coefficients at different frequencies are multiplied by their corresponding Fourier basis functions and are then linearly summed up. The Fourier-based positional encoding plays a similar role as the Fourier basis function here. It greatly enhances the expressivity of $\mathbf{G}_1\big(\mathbf{u}_1^{n}\big)\mathbf{z}^{n}$ in approximating the nonlinear composition $\mathbf{v}(\mathbf{x}_i^n,t_n)$, and facilitates the discovery of a compact latent embedding $\mathbf{z}$.

\paragraph{(c) Low-rank approximation of $\mathbf{G}_2$ for scalability.}
When a relatively high latent dimension is needed for embedding high-dimensional complex flow systems, the latent transition matrix $\mathbf{G}_2 \in \mathbb{R}^{d_{\mathbf{z}} \times d_{\mathbf{z}}}$ can be a computational bottleneck of CGKN, as its parameter count scales quadratically with the latent dimension $d_{\mathbf{z}}$. To ensure scalability while retaining expressive capacity, a low-rank approximation of $\mathbf{G}_2$ is adopted:
\[
\mathbf{G}_2 = \mathbf{U}\,\mathrm{diag}(\mathbf{s})\,\mathbf{V}^\top + \mathrm{diag}(\bm{\delta}),
\]
where $\mathbf{U}, \mathbf{V} \in \mathbb{R}^{d_{\mathbf{z}}\times r}$ are orthonormal matrices, $\mathbf{s}\in\mathbb{R}^r$ contains nonnegative singular values, and $\bm{\delta}\in\mathbb{R}^{d_{\mathbf{z}}}$ is a learnable diagonal correction. The effective rank $r \ll d_{\mathbf{z}}$ controls the trade-off between model capacity and efficiency. During training, $\mathbf{U}$ and $\mathbf{V}$ are obtained by QR orthogonalization of unconstrained matrices $\mathbf{U}_{\text{raw}}$ and $\mathbf{V}_{\text{raw}}$, ensuring that $\mathbf{U}^\top \mathbf{U} = \mathbf{V}^\top \mathbf{V} = \mathbf{I}_r$ at each forward pass, while $\mathbf{s}$ is parameterized through a softplus transformation to enforce positivity. The additional diagonal term $\mathrm{diag}(\bm{\delta})$ serves as a residual correction, improving conditioning and allowing independent scaling across latent modes. This parameterization reduces the number of trainable parameters from $\mathcal{O}(d_{\mathbf{z}}^2)$ to $\mathcal{O}(d_{\mathbf{z}}r)$, substantially decreasing both memory and computational costs in the latent update~\eqref{eq:LaCGKN_z}. In practice, we find that moderate ranks (e.g., $r=64$ to $128$) already capture the essential coupling structures of the latent flow dynamics while maintaining numerical stability. This low-rank SVD-inspired construction of $\mathbf{G}_2$ thus enables LaCGKN to scale efficiently to high-dimensional turbulent flows without compromising much representational power.

\paragraph{Learning LaCGKN from data.}
The LaCGKN is trained by minimizing a composite loss function:
\begin{equation}\label{eq:loss}
L(\boldsymbol{\theta}_{\mathcal{E}},\boldsymbol{\theta}_{\mathcal{D}}, \boldsymbol{\theta}_{\eta}) := \lambda_\textrm{AE} L_{\textrm{AE}}(\boldsymbol{\theta}_{\mathcal{E}},\boldsymbol{\theta}_{\mathcal{D}}) + \lambda_{\mathbf{u}} L_{\mathbf{u}}(\boldsymbol{\theta}_{\mathcal{E}},\boldsymbol{\theta}_{\mathcal{D}}, \boldsymbol{\theta}_{\eta}) + \lambda_{\mathbf{z}} L_{\mathbf{z}}(\boldsymbol{\theta}_{\mathcal{E}}, \boldsymbol{\theta}_{\eta}) + \lambda_{\textrm{DA}}L_{\textrm{DA}}(\boldsymbol{\theta}_{\mathcal{D}}, \boldsymbol{\theta}_{\eta}),
\end{equation}
where
\begin{subequations}
\begin{align}
L_{\textrm{AE}}
&:= \mathbb{E}_{\mathbf{u}_2^\star}
\left[
\frac{1}{d_2}\,
\big\|\mathbf{u}_2^\star
- \mathcal{D}\!\left(\mathcal{E}(\mathbf{u}_2^\star)\right)
\big\|_2^2
\right],
\label{eq:Loss_AE}\\[6pt]
L_{\mathbf{u}}
&:= \mathbb{E}_{\mathbf{u}^{0\star}}
\left[
\frac{1}{N_s}
\sum_{n=1}^{N_s}
\frac{1}{d_\mathbf{u}}\,
\big\|\mathbf{u}^{n\star}-\mathbf{u}^n\big\|_2^2
\right],
\label{eq:Loss_u}\\[6pt]
L_{\mathbf{z}}
&:= \mathbb{E}_{\mathbf{u}^{0\star}}
\left[
\frac{1}{N_s}
\sum_{n=1}^{N_s}
\frac{1}{d_\mathbf{z}}\,
\big\|\mathbf{z}^{n\star}-\mathbf{z}^{n}\big\|_2^2
\right],
\label{eq:Loss_z}\\[6pt]
L_{\textrm{DA}}
&:= \mathbb{E}_{\mathbf{u}^{0\star}}
\left[
\frac{1}{N_l-N_b}
\sum_{n=N_b+1}^{N_l}
\frac{1}{d_2}\,
\big\|\mathbf{u}_2^{n\star}-\boldsymbol{\mu}_2^{n}\big\|_2^2
\right].
\label{eq:Loss_DA}
\end{align}
\end{subequations}
The loss functions \eqref{eq:Loss_AE}--\eqref{eq:Loss_DA} represent the auto-encoder reconstruction loss for Eulerian flow, the forecast loss for physical variables $\mathbf{u}=\{{\mathbf{u}_1}, \mathbf{u}_2\} \in \mathbb{R}^{d_\mathbf{u}}$, the forecast loss for latent variables $\mathbf{z}$, and the data assimilation loss based on the posterior mean $\boldsymbol{\mu}_2^{n}$ obtained from the conditional Gaussian filter and decoder, respectively. The subscript ${}^{\star}$ denotes the reference (ground-truth) solution, and $\|\cdot\|_2$ denotes the Euclidean $\ell^2$-norm. The forecast steps $N_s$, data assimilation steps $N_l$, data assimilation warm-up steps $N_b$, and the loss weights $\lambda_{\textrm{AE}}$, $\lambda_{\mathbf{u}}$, $\lambda_{\mathbf{z}}$, and $\lambda_{\textrm{DA}}$ are tunable hyper-parameters.

To learn the LaCGKN \eqref{eq:LaCGKN} from data, we follow the two-stage training strategy developed for discrete-time CGKN as in \cite{chen_modeling_2025}. In Stage~1, the autoencoder $(\mathcal{E},\mathcal{D})$ and the coefficients $(\mathbf{F}_1,\mathbf{G}_1,\mathbf{F}_2,\mathbf{G}_2)$ are jointly learned by minimizing the composite loss \eqref{eq:loss} with data assimilation loss excluded ($\lambda_{\text{DA}}=0$). Once finishing Stage 1, the noise strength matrix $\boldsymbol{\Sigma}_1$ is estimated by the one-step prediction root mean squared error (RMSE) of the observed states $\mathbf{u}_1$. The noise strength matrix $\boldsymbol{\Sigma}_2$ is manually set in this work, although it can be estimated by the one-step prediction RMSE of latent states $\mathbf{z}$ or set as a trainable parameter as well. In Stage~2, the LaCGKN surrogate is refined for data assimilation by minimizing the complete composite loss \eqref{eq:loss} with $L_{\text{DA}}$ included. The resulting LaCGKN thus learns a model both for tracer--flow prediction and Lagrangian data assimilation.

\begin{algorithm}[H]
\caption{Learning LaCGKN from data}
\label{alg:LaCGKN}
\begin{algorithmic}
    \State \textbf{Input:}
    $\big\{\mathbf{u}_1^{n\star},\mathbf{u}_2^{n\star}\big\}_{n=0}^{N}$ \Comment{Training data: tracer positions and flow fields}
    \State $\boldsymbol{\theta}_{\mathcal{E}}^{(1)},\boldsymbol{\theta}_{\mathcal{D}}^{(1)},\boldsymbol{\theta}_{\eta}^{(1)}
    \leftarrow \arg\min\; \{\lambda_{\textrm{AE}}L_{\textrm{AE}}+\lambda_{\mathbf{u}}L_{\mathbf{u}}+\lambda_{\mathbf{z}}L_{\mathbf{z}}\}$
    \Comment{Train LaCGKN without data assimilation loss}
    \State $\mathrm{diag}(\boldsymbol{\Sigma}_1) \leftarrow \textrm{RMSE}(\mathbf{u}_1^{\star}, \mathbf{u}_1)$, $\mathrm{diag}(\boldsymbol{\Sigma}_2) \leftarrow$ manually set \Comment{Uncertainty quantification for state forecast}
    \State $\boldsymbol{\theta}_{\mathcal{E}}^{(2)},\boldsymbol{\theta}_{\mathcal{D}}^{(2)},\boldsymbol{\theta}_{\eta}^{(2)}
    \leftarrow \arg\min\; \{\lambda_{\textrm{AE}}L_{\textrm{AE}}+\lambda_{\mathbf{u}}L_{\mathbf{u}}+\lambda_{\mathbf{z}}L_{\mathbf{z}}+\lambda_{\textrm{DA}}L_{\textrm{DA}}\}$
    \Comment{Train LaCGKN with data assimilation loss}
    \State $\boldsymbol{\theta}^*_{\mathcal{U}} = \arg\min  \textrm{MSE}\big(\|\mathbf{u}_2^\star - \boldsymbol{\mu}_2\|, \mathcal{U}(\mathbf{u}_1;\boldsymbol{\theta}_{\mathcal{U}})\big)$ \Comment{Uncertainty quantification for data assimilation}
    \State \textbf{Output}: $\boldsymbol{\theta}_{\mathcal{E}}^{(2)}, \boldsymbol{\theta}_{\mathcal{D}}^{(2)}, \boldsymbol{\theta}_{\eta}^{(2)}, 
    \boldsymbol{\theta}^*_{\mathcal{U}}, \boldsymbol{\Sigma}_1, \boldsymbol{\Sigma}_2$ \Comment{Trained parameters and noise strengths}
\end{algorithmic}
\end{algorithm}

\section{Numerical experiments}
\subsection{Application to the two-layer quasi-geostrophic flow with surface tracer observations}
The quasi-geostrophic (QG) equations provide a nonlinear yet analytically clean model that captures the essential physics of mid-latitude atmosphere and ocean dynamics, and are widely used to study large-scale geophysical flows \citep{vallis_atmospheric_2017}. In this work, the LaCGKN is applied to a tracer--flow system in which the flow is governed by a two-layer QG model with baroclinic instability \citep{qi_low-dimensional_2016}:
\begin{subequations}\label{eq:qg}
\begin{align}
\frac{\partial q_1}{\partial t} + J(\psi_1, q_1) + \beta \frac{\partial \psi_1}{\partial x} + U_1 \frac{\partial }{\partial x}\nabla^2 \psi_1 + \frac{k_d^2}{2} \left( U_1 \frac{\partial \psi_{2}}{\partial x} - U_{2} \frac{\partial \psi_1}{\partial x} \right) &=  - \nu \Delta^s q_1,    \\
\frac{\partial q_2}{\partial t} + J(\psi_2, q_2) + \beta \frac{\partial \psi_2}{\partial x} + U_2 \frac{\partial }{\partial x}\nabla^2 \psi_2 + \frac{k_d^2}{2} \left( U_2 \frac{\partial \psi_{1}}{\partial x} - U_{1} \frac{\partial \psi_2}{\partial x} \right) &= - \left( U_2 \frac{\partial h}{\partial x} + \kappa \nabla^2 \psi_2 \right) - \nu \Delta^s q_2,
\end{align}
\end{subequations}
where $\psi_\ell(\mathbf{x}, t)\in \mathbb{R}, (\ell=1, 2)$ denotes the $\ell$th-layer stream function defined on $(\mathbf{x},t)\in \Omega \times (0,T]\subset \mathbb{R}^2\times \mathbb{R}$ and is the primary quantity of interest. The potential vorticity disturbances in the upper and lower layers are defined as
\begin{equation}\label{eq:q1q2}
q_1 = \nabla^2 \psi_1 + \frac{k_d^2}{2} (\psi_2 - \psi_1), \quad q_2 = \nabla^2 \psi_2 + \frac{k_d^2}{2} (\psi_1 - \psi_2) + h,
\end{equation}
respectively. Here, $h$ is the bottom topography, $J(A,B)=A_xB_y-A_yB_x$ is the Jacobian, $(U_1,U_2)$ are the mean flows in the two layers, $k_d$ is the deformation wavenumber associated with the Rossby deformation radius, and $\beta$ is the Rossby parameter. The term $\kappa \nabla^2 \psi_2$ represents Ekman damping in the lower layer (bottom friction), and $ \nu \Delta^s q_i$ denotes hyperviscosity in each layer.  In this work, we set $(U_1,U_2)=(U,-U)$ to impose symmetric vertical shear which introduces baroclinic instability, and consider equal layer depths $H_1=H_2=H/2$. The incompressible velocity field of the $\ell$th layer is related to the stream function by
\begin{equation}\label{eq:v_psi}
\mathbf{v}_\ell=\left(\frac{\partial \psi_\ell}{\partial y},\,-\frac{\partial \psi_\ell}{\partial x}\right)^\top.
\end{equation}
Tracers are treated as passive and are advected by the surface velocity field according to \eqref{eq:tracer_discrete}. 

The two-layer QG flow with surface tracer observations provides a challenging testbed for the LaCGKN due to its intrinsic multiscale turbulent dynamics, strong nonlinearity, and the existence of an unobserved bottom topography. As in \eqref{eq:u1u2_LaCGKN}, LaCGKN takes tracer positions $\big(\mathbf{x}_1^{n},\ldots,\mathbf{x}_I^{n}\big)$ as the observed state $\mathbf{u}_1^{n}$, but the stream function fields $(\psi_1^n,\psi_2^n)$ (rather than the velocity fields) as the unobserved state $\mathbf{u}_2^{n}$. The unobserved state $\mathbf{u}_2^n$ is encoded by a convolutional autoencoder into a latent state $\mathbf{z}^{n}\in\mathbb{R}^{d_{\mathbf{z}}}$ with spatial structure $(z_h,z_w,n_c)$. We find that choosing the number of channels $n_c$
to match the number of layers often yields a better latent representation. Since the domain is doubly periodic, tracer positions are encoded via an angular embedding $(x,y)\ \mapsto\ (\cos x,\ \sin x,\ \cos y,\ \sin y)$, and the convolutional autoencoder uses circular padding in order to respect the periodicity (see Appendix \hyperref[app:A]{A} for details). As described in Section~\ref{sec:LaCGKN}, the LaCGKN observation operators $(\mathbf{F}_1,\mathbf{G}_1)$ employ Fourier positional encoding and share parameters across all tracers. When using a relatively high latent dimension (e.g., $z_h=z_w=32$), the latent evolution matrix $\mathbf{G}_2$ is parameterized by the SVD-inspired low-rank approximation to reduce trainable parameters.

\subsection{Experimental design}
The proposed LaCGKN is evaluated on the two-layer QG flow with surface tracer observations. Its performance in state forecasting and data assimilation is compared against several baseline methods listed below. 

\paragraph{State forecast.} The following methods are evaluated for one-step prediction of both the flow field and tracer positions.
\begin{enumerate}[(i)]
    \item \textit{LaCGKN.} The proposed deep learning method based on the LaCGKN surrogate \eqref{eq:LaCGKN}, which jointly predicts tracer and flow dynamics. The latent dimension is $d_{\mathbf{z}}=16\times16\times2$ by default.

    \item \textit{LaCGKN$_\mathit{32}$.} A higher-capacity variant of LaCGKN with an increased latent dimension $d_{\mathbf{z}} = 32 \times 32 \times 2$, in order to enhance model expressivity.

    \item \textit{DNN(tracer) + CNN(flow).} A modular deep-learning baseline that predicts tracer positions with deep neural network (DNN) and predicts flow with CNN. For tracer prediction, each tracer position is first embedded through the Fourier positional encoding \eqref{eq:FourierEncoding}. Feature-wise Linear Modulation \citep[FiLM;][]{perez_film_2017} is then applied to enhance the retrieval of local flow information at tracer positions:
        \begin{equation}\label{eq:tracer_DNN}
            \mathbf{u}_1^{n+1}
            = \mathbf{u}_1^{n} + \mathrm{DNN}\!\left(\beta(\tilde{\mathbf{u}}^{n}_1)
              + \gamma(\tilde{\mathbf{u}}^{n}_1)\odot \mathbf{u}_2^{n}\right),
        \end{equation}
where $\tilde{\mathbf{u}}^{n}_1=\mathcal{P}(\mathbf{u}_1^{n})$ is the positionally encoded tracer state and $\odot$ is the elementwise product. The modulation parameters $\beta$ and $\gamma$ are given by two shallow fully connected networks. For flow prediction, a CNN with circular padding is used to predict $\mathbf{u}_2^{n+1}$ from $\mathbf{u}_2^{n}$. This modular architecture serves as a strong deep-learning baseline for pure state forecast.
        
    \item \textit{Persistence.} A trivial baseline that assumes no temporal evolution, i.e., $\mathbf{u}^{n+1}=\mathbf{u}^n$. Persistence provides a lower bound on forecast skill and reveals the improvement over simply copying the current state.
\end{enumerate}

\paragraph{Data assimilation.}
The following methods are evaluated for data assimilation, which aims to infer the Eulerian flow state from surface tracer observations.
\begin{enumerate}[(i)]
    \item \textit{LaCGKN.}
    The proposed method, which performs analytic filtering in latent space using the conditional Gaussian filter \eqref{eq:CGFilter}, followed by decoding to physical space. No access to the physical model is used during assimilation.

    \item \textit{Ensemble Kalman filter (EnKF).}
    A classical sequential data assimilation method for nonlinear systems \citep{evensen_sequential_1994}. We employ a deterministic EnKF variant, the ensemble adjustment Kalman filter \citep[EAKF;][]{anderson_ensemble_2001}, using an augmented state formulation for Lagrangian data assimilation as in \cite{wang_nonlinear_2025}.  A constant multiplicative covariance inflation \citep{anderson_monte_1999} and Gaspari--Cohn localization \citep{gaspari_construction_1999} are applied. EAKF runs ensemble forecast using the perfect QG model. It serves as a strong physics-based baseline and highlights the accuracy and efficiency of LaCGKN.

    \item \textit{Optimal Interpolation (OI).}
    A simplified Kalman-filter-based method with prescribed, time-invariant background and observation error covariances. Following \cite{molcard_assimilation_2003,molcard_lagrangian_2005}, OI is applied by constructing Lagrangian velocities via finite-differencing tracer positions over $\Delta t_{\mathrm{obs}}$ (thus converting position observations into velocity observations), and projecting surface-layer corrections to the lower layer via linear regression. Gaussian localization is applied. OI runs the flow forecast with the perfect QG model. It provides a simplified baseline and contrasts the importance of capturing strong nonlinearity in tracer–flow dynamics.

    \item \textit{Climatology.}
    A naive baseline that ignores observations and uses the long-term mean flow as a fixed background state. It provides a reference error level without assimilation.
\end{enumerate}

\paragraph{Data generation.}The tracer-flow system is integrated numerically to generate paired Lagrangian tracer observations and Eulerian QG flow fields. The QG flow evolves in a turbulent regime with nondimensional parameters $k_d=10$, $\beta=22$, $U=1$, $\kappa=9$, $\nu=10^{-12}$, $s=4$, and $h(x,y)=40\cos x + 80\cos(2y)$. The QG equations are solved on a $128\times128$ pseudo-spectral grid over the doubly periodic domain $\Omega=[0,2\pi)^2$. The flow is initialized by sampling the potential vorticity disturbances as $q_1^0=q_2^0 \sim \,\mathcal{N}(0,10^2)$ and integrated for $N_t=2\times10^{6}$ steps after a warm-up of $1000$ steps, with time step $\Delta t=2\times10^{-3}$. A total of 1024 tracers are initialized as $\mathbf{x}_i^0 \sim \mathcal{N}(\pi,0.1^2)$ and advected by the surface flow with stochastic noises of strength $\sigma_x=\sigma_y=0.1$. Synthetic tracer observations are constructed by adding independent Gaussian measurement noise $\mathcal{N}(0,0.01^2)$. The same level of noise is added to the stream function fields during training to mimic realistic deployments and assess robustness. The data are then sub-sampled to a $64\times64$ grid with observation interval $\Delta t_{\mathrm{obs}}=4\times10^{-2}$. The resulting dataset consists of tracer position data $\{\mathbf{x}_i^n\}_{i=1,n=0}^{I,N_{t_{obs}}}$ and Eulerian QG flow fields $\{\psi_{1,m}^n, \psi_{2,m}^n\}_{m=1,n=0}^{M,N_{t_{obs}}}$, where $M=64\times 64$ and $N_{t_{obs}}=10^5$. The dataset is split into training/validation/testing segments with $80{,}000/10{,}000/10{,}000$ assimilation steps, respectively. Unless otherwise stated, only $I=64$ tracers are used in forecast and assimilation. During training, a random subset of 64 tracers is drawn from the total $1024$ tracers at each batch to encourage tracer homogenization of ($\mathbf{F}_1,\mathbf{G}_1)$ and improve generalization.

\paragraph{LaCGKN training setup.}
The LaCGKN consists of a circular CNN autoencoder $(\mathcal{E},\mathcal{D})$ for the Eulerian flow and a conditional Gaussian network \eqref{eq:LaCGKN} with coefficients $(\mathbf{F}_1,\mathbf{G}_1,\mathbf{F}_2,\mathbf{G}_2)$. By default, the encoder $\mathcal{E}:\mathbb{R}^{64\times64\times2}\mapsto\mathbb{R}^{16\times16\times2}$ maps the flow field $\mathbf{u}_2^n$ to a latent state $\mathbf{z}^n$ ($d_{\mathbf{z}}=512$). The decoder $\mathcal{D}:\mathbb{R}^{16 \times 16 \times2}\mapsto\mathbb{R}^{64 \times 64\times 2}$ mirrors the encoder, mapping $\mathbf{z}^n$ back to $\mathbf{u}_2^n$. Given angular embedded tracer positions $\mathbf{u}_1^n \in \mathbb{R}^{4I}$, each angular representation $(\cos x_i,\sin x_i,\cos y_i,\sin y_i)$ is first converted  back to $(x_i,y_i)$ and then embedded via Fourier positional encoding with $K=6$ frequencies, yielding $\tilde{\mathbf{x}}_i^{n}=\mathcal{P}(\mathbf{x}_i^{n}) \in \mathbb{R}^{26}$ per tracer. The sub-networks $\mathbf{f}_\theta:\mathbb{R}^{26}\to\mathbb{R}^{4}$ and $\mathbf{G}_\theta:\mathbb{R}^{26}\to\mathbb{R}^{4\times 512}$ that are DNNs take $\tilde{\mathbf{x}}_i^{n}$ as input, and construct $\mathbf{F}_1\in\mathbb{R}^{4I}$ and $\mathbf{G}_1\in\mathbb{R}^{4I\times 512}$. The latent transition parameters $\mathbf{F}_2\in\mathbb{R}^{512}$ and $\mathbf{G}_2\in\mathbb{R}^{512\times 512}$ are learned constants. To explore higher-capacity models, a larger latent dimension $d_{\mathbf{z}}=32\times32\times2$ is also considered, referred to as LaCGKN$_{32}$; in this case, an SVD-inspired low-rank parameterization of $\mathbf{G}_2$ (Section~\ref{sec:LaCGKN}) is applied with effective rank $r=64$. The training settings of LaCGKN are:  state forecast steps $N_s=1$ (i.e., one-step prediction), data assimilation steps $N_l=100$, data assimilation warm-up steps $N_b=20$, and the loss weights $\lambda_{\textrm{AE}}=\lambda_{\mathbf{u}}=\lambda_{\mathbf{z}}=\lambda_{\textrm{DA}}=1$.

\paragraph{Test and evaluation.}
Over the $10{,}000$-step testing dataset, the one-step prediction error is evaluated for state forecast, and posterior error is evaluated for data assimilation with the first 50 steps excluded as spin-up. LaCGKN performs analytic latent-space filtering initialized as
$\mathbf{z}^0\sim\mathcal{N}(\mathbf{0},0.1^2\mathbf{I})$. EnKF and OI perform sequential filtering using the perfect QG model, with flow initialized as $\psi_{\ell, m}^{0} \sim \mathcal{N}(0,0.1^2)$ at each grid point and in each layer. For EnKF, each simulated tracer is initialized around the initial tracer observation $\mathbf{x}^{0} \sim\mathcal{N}(\mathbf{x}_{\mathrm{obs}}^{0},0.01^2\mathbf{I})$. EnKF uses ensemble size $N_e=40$, inflation factor 1.025, and localization radius of 16 grid points. OI uses a prescribed background error standard deviation $\sigma_{b}=1.0$ of velocity and Gaussian localization radius of 1 grid point. The regression coefficients used for projecting surface corrections to the lower layer is calculated based on per-grid velocities between the two layers. All tunable parameters are empirically optimized.

\paragraph{Evaluation metric.}
The RMSE is used as the point-estimation metric. For a quantity of interest $\mathbf{a} \in \mathbb{R}^{d_{\mathbf{a}}}$ (e.g., stacked tracer positions or a flattened flow field) and its reference value ${\mathbf{a}}^{\star}$, the RMSE is
\begin{equation}\label{eq:rmse_inst}
\mathrm{RMSE}
:=\left(\frac{1}{d_{\mathbf{a}}}\,\big\|\mathbf{a}-{\mathbf{a}}^{\star}\big\|_2^2\right)^{1/2}.
\end{equation}
For data assimilation, EnKF and OI use the posterior mean as the point estimate, while LaCGKN uses the decoded latent posterior mean $\boldsymbol{\mu}_2=\mathcal{D}(\boldsymbol{\mu}_{\mathbf{z}})$. For all deep-learning-based methods, the number of trainable parameters is chosen to be comparable to ensure a fair comparison. The trainable parameter counts are $1,021,512$ for the default LaCGKN, $927,932$ for LaCGKN$_{32}$, and $1,089,894$ for DNN+CNN.

\subsection{Numerical results}
\begin{figure}[ht]
 \centerline{\includegraphics[width=33pc]{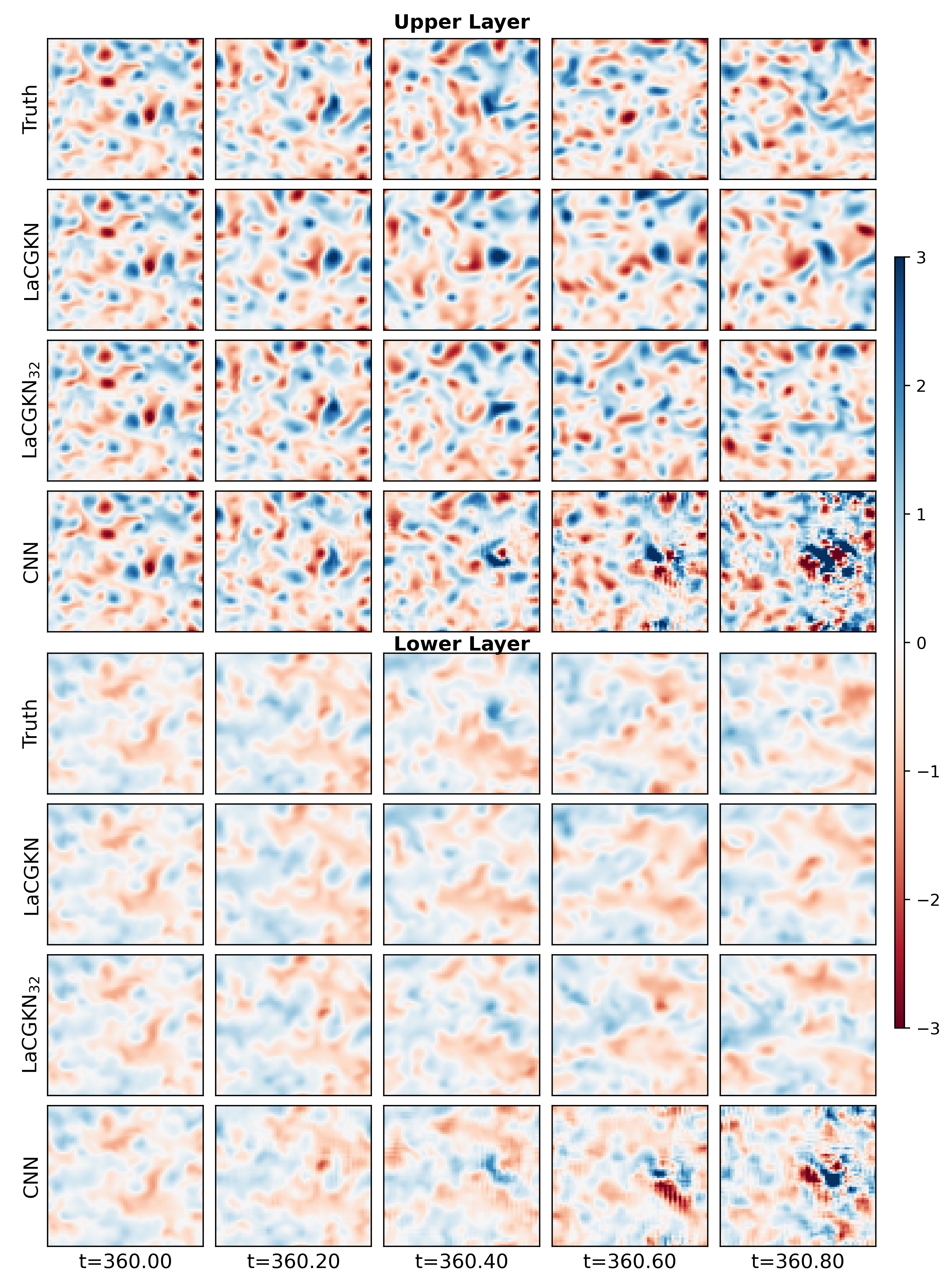}}
    \caption{Snapshots of the stream function flow field for state forecast comparison. The forecasts are initialized at time $t=360$, and are autoregressly rolled out for multiple steps. Results are compared among the ground truth, LaCGKN (default latent dimension $d_{\mathbf{z}}= 16\times 16\times2$), $\text{LaCGKN}_{32}$ (latent dimension $d_{\mathbf{z}}= 32\times 32\times2$) and CNN.}
    \label{fig:flow_comparison_pred}
\end{figure}

\begin{figure}[ht]
 \centerline{\includegraphics[width=33pc]{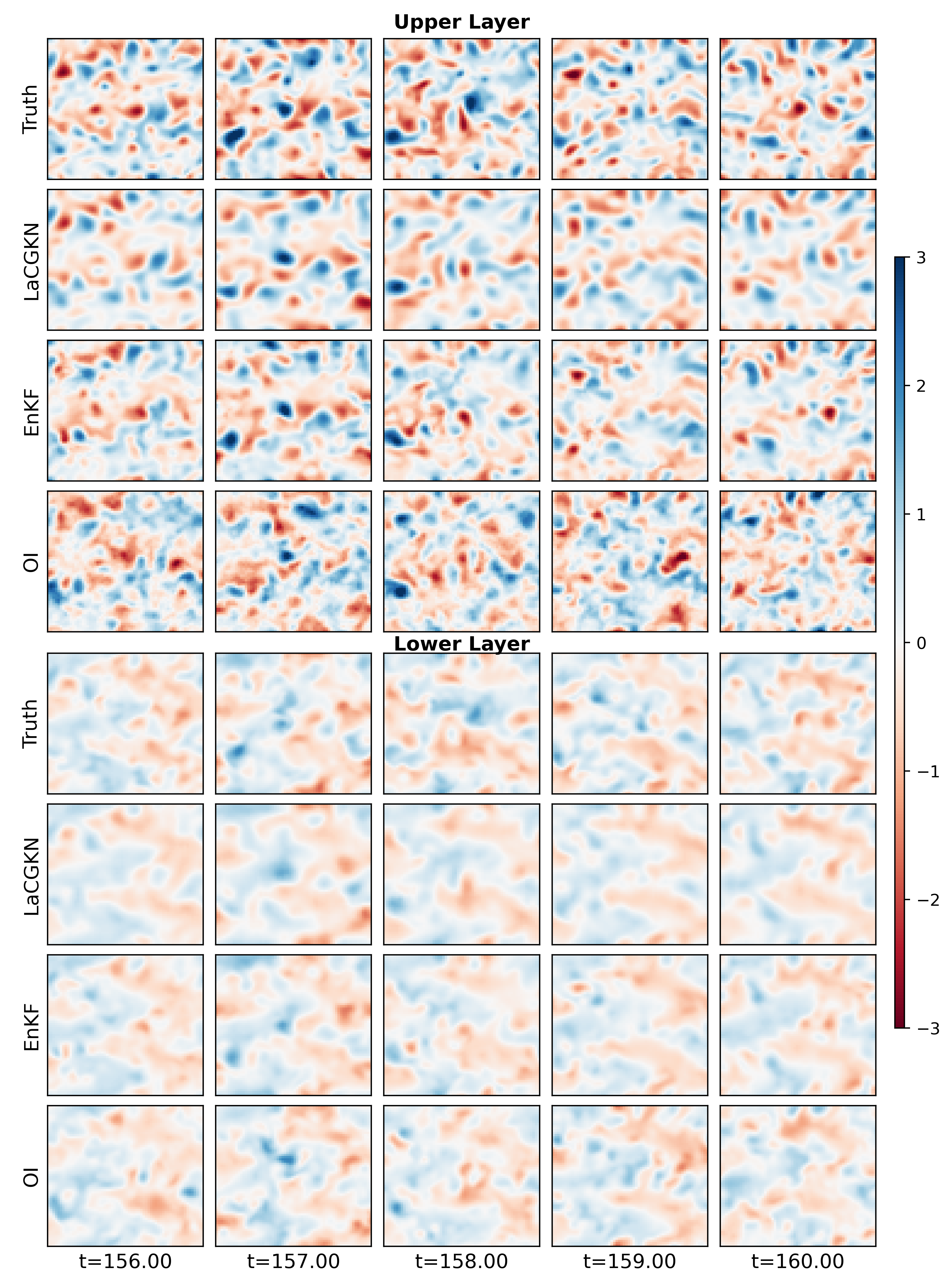}}
    \caption{Snapshots of the stream function flow field for data assimilation comparison. The ground truth is compared to the posterior estimates from LaCGKN, EnKF, and OI.}
    \label{fig:flow_comparison_da}
\end{figure}

\begin{figure}[ht]
 \centerline{\includegraphics[width=35pc]{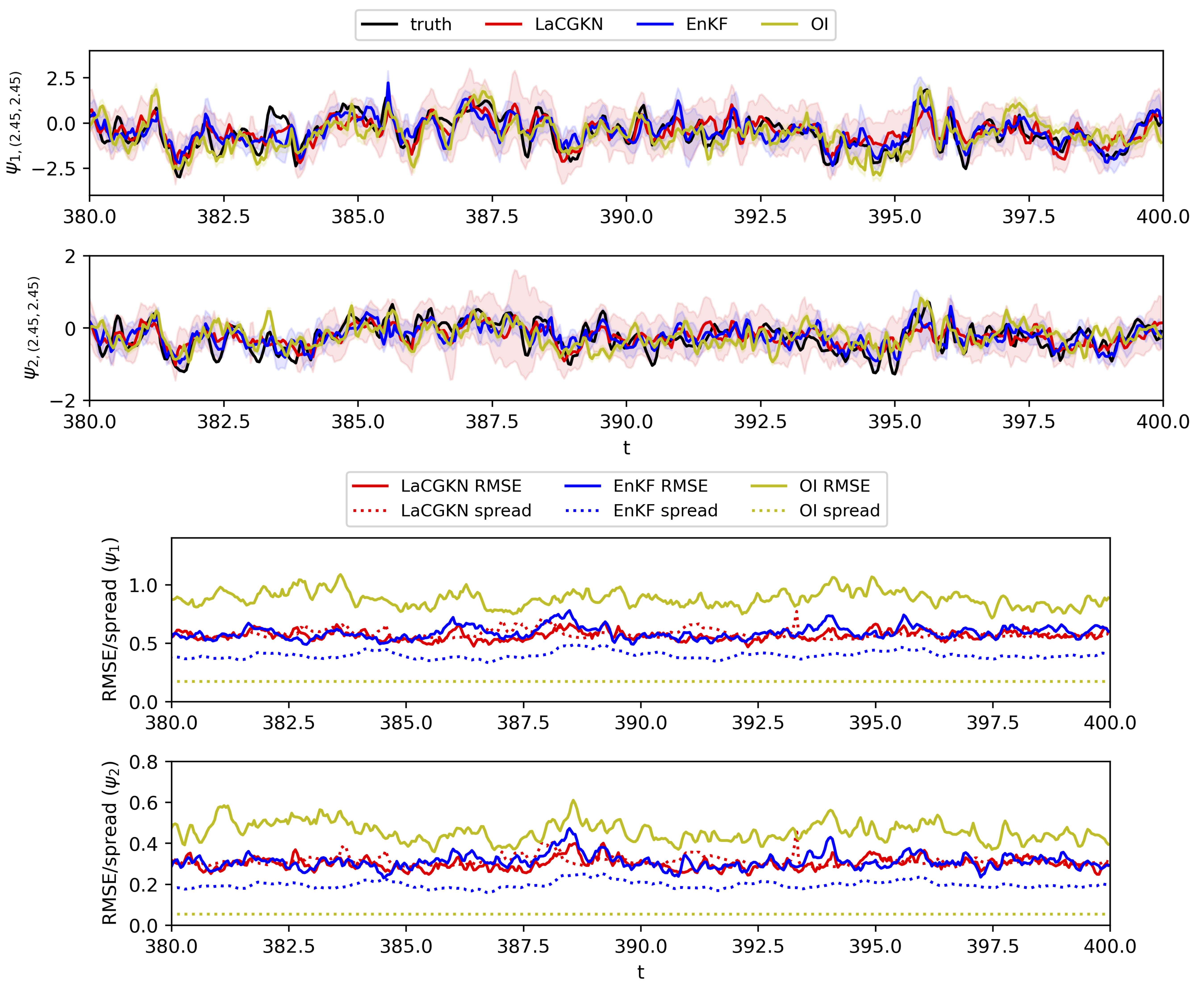}}
  \caption{(a) Time series of the data assimilation posterior estimates at position $(x,y)=(2.45, 2.45)$. Truth is denoted by black lines. The comparing methods are LaCGKN (red lines), EnKF (blue lines), and OI (yellow lines). Uncertainties are illustrated as $\pm 2$ posterior standard deviation using the corresponding shaded areas. (b) Time series of the data assimilation posterior RMSEs (solid lines) and posterior spread (standard deviation; dashed lines) for LaCGKN (red lines), EnKF (blue lines), and OI (yellow lines).}\label{fig:series}
\end{figure}

\begin{figure}[ht]
 \centerline{\includegraphics[width=31pc]{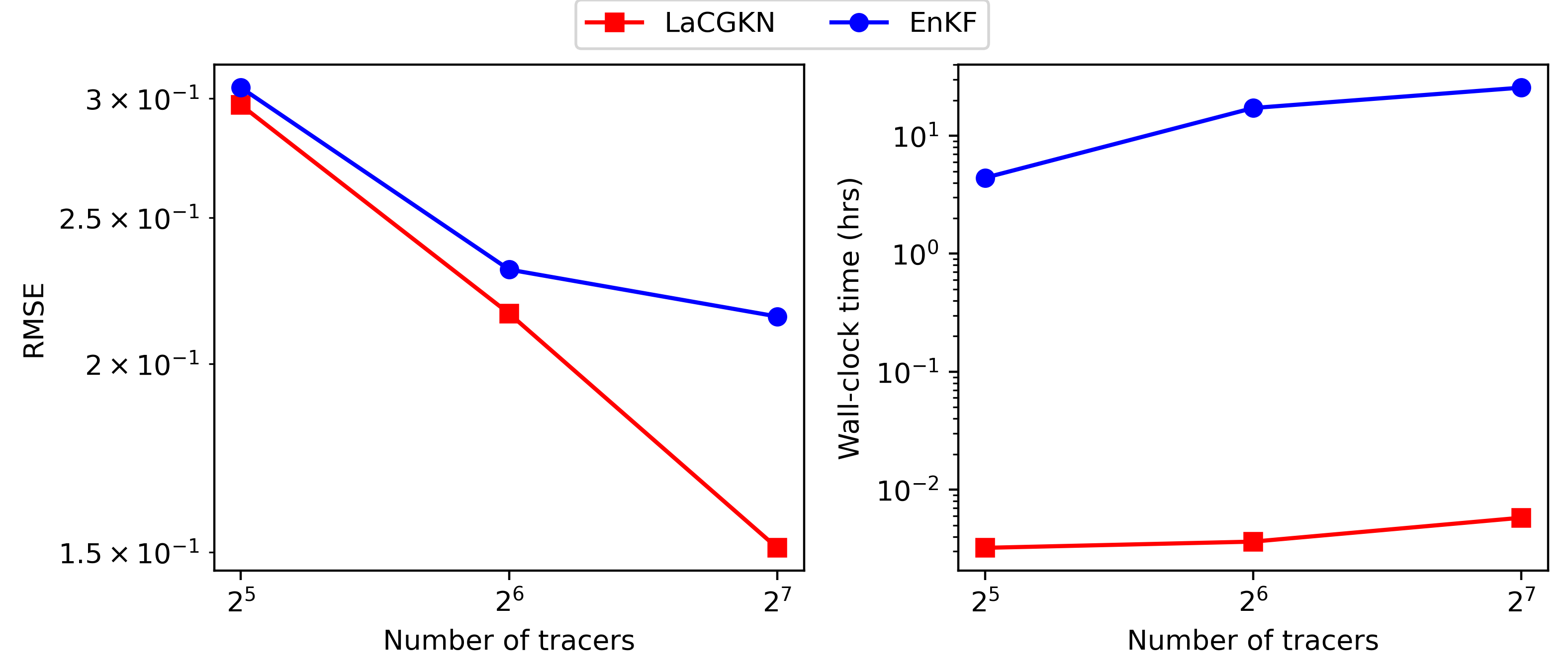}}
  \caption{Posterior RMSEs and computational costs (wall-clock time) of LaCGKN and EnKF varying with number of tracers. The wall-clock time is collected by running the program on the same CPU. The model forecasting of EnKF is parallelized over ensemble members.}\label{fig:sensitivity_cost}
\end{figure}

The performance of LaCGKN is evaluated in terms of both one-step prediction and data assimilation. A key distinction is that LaCGKN performs forecasting and data assimilation within a unified framework, whereas the benchmark methods considered below address only one of these two tasks. The key findings are summarized below:
\begin{itemize}
    \item \textbf{Accurate and stable flow forecasting.}
     The default LaCGKN substantially outperforms persistence, demonstrating its ability to learn nontrivial dynamical structure. Increasing the latent dimension ($\mathrm{LaCGKN}_{32}$) yields the lowest RMSE among all comparing methods and maintains stable long-term rollouts, whereas the CNN baseline becomes unstable after about 10 steps (Table~\ref{tab:rmse_pred} and Fig.~\ref{fig:flow_comparison_pred}).

    \item \textbf{Accurate data assimilation with reliable uncertainty quantification.}
    Without access to the physical model, LaCGKN achieves the lowest posterior RMSEs with coherent flow structures compared to the physics-based EnKF, OI and climatology that use the perfect QG model (Table~\ref{tab:rmse_da} and Fig.~\ref{fig:flow_comparison_da}). LaCGKN also provides well-calibrated posterior uncertainty quantification, with the posterior spread closely matching the posterior RMSE over time (Fig.~\ref{fig:series}).

    \item \textbf{High computational efficiency and scalability.}
    LaCGKN is approximately two orders of magnitude faster than the parallelized EnKF while achieving lower posterior RMSE, and its efficiency advantage becomes more pronounced as the number of tracers increases. Furthermore, once trained, LaCGKN generalizes across different tracer numbers without retraining (Fig.~\ref{fig:sensitivity_cost}).
\end{itemize}

The detailed results are presented as follows.
\paragraph{State forecast.}
The one-step prediction RMSEs across different methods are summarized in Table \ref{tab:rmse_pred}. Both the default LaCGKN and the LaCGKN$_{32}$ have substantially lower RMSEs than persistence across all variables, indicating that the learned surrogates capture meaningful dynamical structure rather than merely memorizing the data. The improvement is especially pronounced for the upper-layer flow, where persistence errors are large due to strong turbulence variability. Although the default LaCGKN has larger RMSEs than the DNN+CNN baseline, increasing the latent dimension of LaCGKN significantly improves the flow prediction accuracy. In particular, LaCGKN$_{32}$ reduces the two-layer RMSE from 0.104 to 0.037, achieving the lowest RMSEs among all methods. This substantial gain highlights the importance of latent embedding capacity in representing complex turbulent flows. When sufficient latent dimension is provided, the conditional Gaussian surrogate can well approximate the underlying nonlinear dynamical system. 

For tracer position prediction, LaCGKN does not outperform DNN+CNN. This is expected since the DNN baseline directly exploits the flow fields in physical space to advect tracers, whereas LaCGKN relies on latent embeddings that are required to capture both the tracer-flow interactions and the flow dynamics. Nevertheless, the primary focus of this work is on accurate flow prediction, where LaCGKN$_{32}$ clearly excels. This advantage is further illustrated in the autoregressive multi-step prediction of the flow field (Fig. \ref{fig:flow_comparison_pred}), where LaCGKN$_{32}$ maintains stable and physically coherent flow structures over long rollouts, while the CNN model becomes unstable after about 10 steps. These results highlight that LaCGKN is not merely a predictor but a dynamical surrogate enforcing structured latent evolution, which encourages long-horizon stability. Moreover, LaCGKN can perform efficient data assimilation beyond standalone forecasting, providing a key advantage over purely predictive deep learning models. 

 \begin{table}[htbp]
    \centering
    \caption{RMSEs of state forecast (one-step prediction) across different methods, including LaCGKN with default latent dimension, LaCGKN$_{32}$ with an increased latent dimension $d_{\mathbf{z}} = 32 \times 32 \times 2$, DNN(tracer)+CNN(flow), and persistence. Errors are computed for tracer positions, the upper-layer stream function $\psi_1$, the lower-layer stream function $\psi_2$, and the two-layer stream functions in total.}
    \begin{tabular}{lcccc}
        \hline
        \textbf{Method} & \textbf{Tracer} & \textbf{Upper Layer} & \textbf{Lower Layer}  & \textbf{Two Layers} \\
        \hline
        LaCGKN & 0.099 & 0.125  & 0.079 & 0.104  \\
        LaCGKN$_{32}$ & 0.094 & 0.042  & 0.032 & 0.037  \\
        DNN+CNN & 0.064  & 0.071 & 0.069 & 0.070 \\
        Persistence & 0.136 & 0.294  & 0.177 & 0.243  \\        
        \hline
    \end{tabular}
    \label{tab:rmse_pred}
\end{table}

\paragraph{Data assimilation.} 
Table \ref{tab:rmse_da} summarizes the posterior RMSEs of flow estimates across different methods. Without using the physical QG model, the default LaCGKN achieves lower RMSEs than EnKF, OI, and climatology in both layers. The advantage of LaCGKN over OI is particularly pronounced. Because OI only introduces local corrections near tracer locations and propagates corrections through linear projections between layers, it performs poorly in this strongly nonlinear turbulent regime. EnKF as a stronger physics-based benchmark more effectively captures nonlinearity by propagating ensemble forecasts with full physical model and estimating flow-dependent error covariances. Nevertheless, LaCGKN achieves lower posterior RMSEs than EnKF without requiring full-model ensemble propagation or empirical tuning of hyper-parameters such as inflation and localization. Snapshots of the flow fields in Fig. \ref{fig:flow_comparison_da} further confirm that LaCGKN accurately recovers coherent flow structures in both layers. These results indicate that the learned LaCGKN surrogate successfully captures the nonlinear tracer-flow coupling and multi-layer interactions required for accurate data assimilation. In addition, the close agreement between posterior spread and the posterior RMSE over time demonstrates that LaCGKN provides reliable uncertainty quantification alongside accurate mean estimates (Fig. \ref{fig:series}). 

\paragraph{Sensitivity and computational cost.}
Figure \ref{fig:sensitivity_cost} compares the posterior RMSEs and computational costs (wall-clock time on a CPU) of LaCGKN and EnKF under varying numbers of tracers. The results admit several key observations. First, LaCGKN consistently achieves lower posterior RMSEs than EnKF across tracer numbers. The performance gap widens as the number of tracers increases, suggesting that the learned surrogate effectively leverages additional observational information. Second, LaCGKN is approximately $O(100)$ times faster than the parallelized EnKF. This efficiency gain arises from filtering in a low-dimensional latent space using closed-form updates, rather than propagating and updating ensembles in the full physical space. The computational complexity scales primarily with latent dimension rather than ensemble size. Third, LaCGKN is invariant to tracer numbers at inference time. Once trained for a given configuration, the same model can be directly applied to different tracer numbers without retraining. In contrast, EnKF requires retuning of hyper-parameters to maintain optimal performance as the number of tracers changes. This tracer-number invariance greatly reduces the operational cost in realistic deployments, where the number of drifting instruments may vary over time. Overall, these results demonstrate that LaCGKN not only performs accurate data assimilation with reliable uncertainty quantification, but also remains highly efficient in both training and inference.

\begin{table}[htbp]
    \centering
    \caption{RMSEs of data assimilation posterior estimates across different methods, including LaCGKN with default latent dimension, EnKF, OI, and climatology. Errors are computed for the upper-layer stream function $\psi_1$, the lower-layer stream function $\psi_2$, and the two-layer stream functions in total.}
    \begin{tabular}{lccc}
        \hline
        \textbf{Method} & \textbf{Upper Layer} & \textbf{Lower Layer} & \textbf{Two Layers} \\
        \hline
        LaCGKN & 0.579  & 0.310 & 0.464  \\
        EnKF  & 0.599  & 0.321 & 0.481  \\
        OI  & 0.890  & 0.467 & 0.710  \\
        Climatology & 0.870  & 0.414 & 0.681  \\
        \hline
    \end{tabular}
    \label{tab:rmse_da}
\end{table}

\section{Conclusion and discussions}
A Lagrangian Conditional Gaussian Koopman network (LaCGKN) is developed to address the challenging tasks of Lagrangian data assimilation and prediction, focusing on inferring high-dimensional Eulerian flow fields from Lagrangian tracer observations. The main difficulty stems from the intrinsic nonlinearity induced by evaluating the flow at moving tracer positions, together with the strongly nonlinear and turbulent nature of flow dynamics. LaCGKN addresses this challenge by learning a surrogate model with a conditional Gaussian structure: the Eulerian flow is encoded into a low-dimensional latent state with approximately linear dynamics, while represented by a nonlinear yet structured model driven by the latent flow. This structure enables analytic latent-space filtering with closed-form posterior updates. To improve robustness and scalability in the Lagrangian setting, we introduce (i) tracer homogenization to enforce permutation equivariance and enable generalization across tracer counts, (ii) Fourier-based positional encoding to capture rich nonlinear spatial dependence of the flow at tracer locations, and (iii) an SVD-inspired low-rank parameterization of the latent transition matrix $\mathbf{G}_2$ to reduce the number of trainable parameters while retaining expressivity and efficiency. The resulting LaCGKN is trained via a composite loss, yielding a unified deep learning model for both prediction and data assimilation.

 The LaCGKN is evaluated on a turbulent two-layer quasi-geostrophic (QG) system with surface tracer position observations. For state forecast, the default LaCGKN already provides a useful forecast model by substantially outperforming persistence, and increasing the latent dimension (LaCGKN$_{32}$) yields the best flow prediction accuracy among the comparing methods. LaCGKN also produces stable and accurate multi-step rollouts, whereas the CNN baseline becomes unstable after about 10 steps. For data assimilation, LaCGKN achieves posterior RMSEs that are comparable to a physics-based EnKF that uses the perfect QG model and significantly lower than those of OI and climatology, while also providing reliable uncertainty quantification. In addition, LaCGKN is very computationally efficient, approximately $O(100)$ times faster than the parallelized EnKF, while maintaining strong accuracy, with the advantages becoming more pronounced as the number of tracers increases. Finally, because LaCGKN is designed to be invariant to tracer number, it can be trained once and applied to different tracer numbers without retraining, reducing both development and deployment costs and providing an efficient, purely data-driven alternative for Lagrangian data assimilation and prediction.

The conditional Gaussian (CG) filter generalizes the Kalman-Bucy filter to systems with nonlinear observation processes, originating from the continuous-time formulation of filtering \citep{kalman_new_1961, jazwinski_5_1970, chen_conditional_2018}. While the discrete-time formulation of filtering, typified by the Kalman filter, is fundamentally different from its continuous-time counterpart \citep{wang_nonlinear_2025}. The distinction is not merely due to time discretization, as discrete-time versions of CG filter also exist, but rather arises from the observation formulation itself. In standard discrete-time filtering, observation noise is independent of system dynamics, whereas in continuous-time filtering (and its discretized surrogates), observation noise enters directly into the dynamics.  In the current CGKN formulation,  measurement errors in tracer positions are not explicitly modeled but are implicitly approximated  by dynamical noise. Although our numerical results demonstrate robustness to small measurement noise, it is of interest to explicitly incorporate measurement noise into the CGKN framework, potentially bridging continuous- and discrete-time filtering formulations, and further improving assimilation accuracy under noisy observations.

\clearpage
\paragraph*{Acknowledgments.}
The research of N.C. is funded by the Office of Naval Research N00014-24-1-2244 and the Army Research Office W911NF-23-1-0118. Z.W. is partially supported by these grants. The research of C.C. and J.W. was funded by the University of Wisconsin-Madison, Office of the Vice Chancellor for Research and Graduate Education, with funding from the Wisconsin Alumni Research Foundation.

%
%
\paragraph*{Data availability statement.}
The code can be found at:
\url{https://github.com/zhongruiw/LaCGKN}.

\section*{Appendix A. Technical details for handling periodic conditions.}\label{app:A}

\subsection*{A.1. Circularity of tracer positions.}\label{app:A1}
    Tracers evolve on a doubly-periodic domain, and their raw Cartesian coordinates exhibit discontinuities upon crossing boundaries. This makes standard Euclidean representations unsuitable, as a small displacement across the $2\pi$-boundary would appear as a large jump. To avoid such artifacts, all tracer positions are encoded using an angular embedding
    \[
        (x,y)\ \mapsto\ (\cos x,\ \sin x,\ \cos y,\ \sin y),
    \]
    which preserves periodicity and continuity on the torus. This unit-vector representation is then mapped back to physical coordinates and further augmented with Fourier features, before being fed into the networks $\mathbf{f}_\theta$ and $\mathbf{G}_\theta$. This representation ensures that neural networks are exposed to geometrically consistent inputs and prevents spurious discontinuities in the training data.

\subsection*{A.2. Circular CNN autoencoder for flow.}\label{app:A2}
    The Eulerian flow field on the $64\times 64$ doubly-periodic grid is encoded and decoded by a convolutional autoencoder whose layers all use wrap-around (circular) padding in both spatial directions. Thus, each convolution effectively acts on a torus, so that no artificial edges are created. The encoder $\mathcal{E}$ compresses the multi-channel flow into the latent state $\mathbf{z}^{n}$ via stacked circular $3\times 3$ convolutions and pooling, and the decoder $\mathcal{D}$ inverts this mapping using circular transposed convolutions. This design yields a compact, translation-equivariant latent representation that enforces periodicity of the reconstructed flow by construction.

\bibliographystyle{ametsocV6}
\bibliography{references}

\end{document}